\documentclass[12pt]{article}
\usepackage{amsmath}
\usepackage{amssymb}
\usepackage{amsfonts}
\usepackage{eucal}

\title{\small \bf Note about leptogenesis from gravity waves\\
in models of inflation}
\author{\small David Lyth and Yeinzon Rodr\'iguez$^\dagger$\\
\small Department of Physics, Lancaster University,\\
\small Lancaster LA1 4YB, UK\\
\small $^\dagger$Centro de Investigaciones, Universidad Antonio Nari\~no,\\
\small Cll 58A \# 37-94, Bogot\'a D.C., Colombia\\
\\
\small Carlos Quimbay\footnote{Associate researcher of the Centro
Internacional de F\'isica,
Ciudad Universitaria, Bogot\'a, D. C., Colombia.} \\
\small Departamento de F\'{\i}sica, Universidad Nacional de Colombia,\\
\small Ciudad Universitaria, Bogot\'a D.C., Colombia}
\date{\small June 2004}

\usepackage[activeacute,english]{babel}

\usepackage{latexsym}

\newcommand{\bn}{\begin{eqnarray}}
\newcommand{\en}{\end{eqnarray}}
\newcommand{\be}{\begin{equation}}
\newcommand{\ee}{\end{equation}}
\newcommand{\no}{\nonumber}
\newcommand{\la}{\label}
\newcommand{\re}{\ref}
\newcommand{\ci}{\cite}

\begin{document}
\maketitle
\begin{abstract}
\noindent We show how the mechanism recently proposed by
Alexander, Peskin, and Sheikh-Jabbari, in which the observed
cosmic matter-antimatter asymmetry comes from the contribution of
the cosmological tensor perturbations to the gravitational anomaly
in the Standard Model, does not work. This holds when taking into
account the commutation relations for the creation-annihilation
\linebreak{operators} associated to the gravitational waves and
the geometrical properties of the
polarization tensors. \\
\end{abstract}

Quite recently Alexander, Peskin, and Sheikh-Jabbari \ci{SPS}
described a mechanism to generate the cosmological matter-antimatter
asymmetry from the gravitational anomaly in the Standard Model
\ci{AGW}:

\be \partial_\mu J^\mu_\ell = \frac{1}{16\pi^2} R \tilde R, \la{an1}
\ee where a non-zero value for the gravitational term \be R \tilde R
= {\small \epsilon}^{\alpha \beta \gamma \delta} R_{\alpha \beta
\rho \sigma} R_{\gamma \delta}{}^{\rho \sigma}, \ee could lead to a
non-conserved leptonic current \be J^\mu_\ell = \bar \ell_i
\gamma^\mu \ell_i + \bar \nu_i \gamma^\mu \nu_i. \ee

It was claimed in \ci{SPS} that a contribution to $R\tilde R$ of
definite sign might be generated by the tensor cosmological
perturbations, produced during inflation, if the inflaton field
contained a CP-odd component.  This could be possible if the
inflaton were a complex modulus field and the imaginary part
$\phi$ of this field could couple to gravity through an
interaction of the type \be \Delta \L = F(\phi) R \tilde R \ ,
\la{axco1} \ee being $F$ odd in $\phi$. This kind of interaction
could be the source of the observable parity-violation in the
cosmic microwave background \ci{LWK} or might cause an asymmetric
evolution of the two polarization states of the tensor-type
perturbation as it is shown in \ci{CHH}. A simple form for
$F(\phi)$ with the correct scaling is \ci{SPS}

\be F(\phi) = \frac {1}{16\pi^2 M_{Pl}} {\cal N} \phi \ , \la{Fval}
\ee where the $M_{Pl}$ in the denominator is approximately the
string scale.

To quantitatively estimate the lepton number produced during
inflation \ci{INF}, it is necessary to compute the production of
gravitational waves under the influence of the coupling in Eq.
(\re{axco1}). When the inflaton field has a slowly-rolling nonzero
classical value, the coupling in the Eq. (\re{axco1}) can lead to
quantum fluctuations of the gravitational field that, treated to
second order, generate a nonzero right-hand side for the
expression in Eq. (\re{an1}).

The general form of the cosmological tensor perturbations in a
Friedman-{\linebreak Robertson-Walker} Universe can be
parameterized as \be ds^2 = dt^2 -
a^2(t)\left(\delta_{ij}+h_{ij}\right)dx^idx^j, \ee where $h_{ij}$
denotes the tensor fluctuations of the metric. Thus, the
contribution of the tensor perturbations to $R\tilde R$, up to
second order in $h_{ij}$, is

\bn R \tilde R = &-& \frac{2}{a^3} \, \epsilon^{ijk} \Bigl[ \left(
\frac{\partial^2}{\partial_l \partial t}h_{jm}
\frac{\partial^2}{\partial_m \partial_i}h_{kl} -
\frac{\partial^2}{\partial_l \partial t}h_{jm}
\frac{\partial^2}{\partial_l \partial_i}h_{km} \right) \no \\
 &+& a^2 \left(\frac{\partial^2}{\partial t^2}h_{jl}
\frac{\partial^2}{\partial_i \partial t}h_{lk} \right)
+\frac{1}{2}\frac{\partial}{\partial t}a^2\left(
\frac{\partial}{\partial
t}h_{jl}\frac{\partial^2}{\partial_i\partial t} h_{lk} \right)
\Bigr] \la{rr1}, \en where we have considered gravity waves moving
in an arbitrary direction in the three-space.

By adding the expression in Eq. (\re{axco1}) to the Einstein action
and varying with respect to the metric fluctuations, it is possible
to find the equations of motion

\be M_{Pl}^2 \Box \, h_{ij} = -\frac{4}{a} \, \epsilon^{ilk}
\frac{\partial^2}{\partial_k \partial t}h_{jl} \left( F''
\dot\phi^2 + 2 H F'\dot\phi \right) \la{me1},\ee where we have
dropped terms with third-order derivatives of $h_{ij}$ and
neglected the acceleration of the inflaton field $\ddot \phi$. As
it was described in \ci{LWK} the coupling in Eq. (\re{axco1})
makes the term $R \tilde R$ different to zero by generating a
``cosmological birefringence" during inflation. This cosmological
birefringence assigns different dispersion relations for the
right- and left-handed polarized waves so that the expression in
Eq. (\re{rr1}) does not vanish.

Writing the tensor perturbations $h_{ij}$ in conformal time, and
expanding them in Fourier modes \be h_{ij}({\bf x}, \tau) =
\frac{\sqrt{2}}{M_{Pl}} \int  \frac{d^3 k}{(2 \pi)^{3/2}(2
\omega_k)^{1/2}} \sum_p \left[e^{i {\bf k \cdot \, x}}h(p,{\bf k},
\tau) \epsilon_{ij}(p,{\bf k})a(p,{\bf k}) + h.c. \right]
\la{Fou1}, \ee where the summation on $p$ is over the two
polarization states of the gravitational waves $(+, \times )$, we
can calculate the vacuum expectation value of the gravitational
operator $R \tilde R$ (Eq. (\re{rr1})) written this time in
conformal time: \be R \tilde R = -\frac{2}{a^4} \, \epsilon^{ijk}
\left( \frac{\partial^2}{\partial_l
\partial \tau}h_{jm} \frac{\partial^2}{\partial_m \partial_i}h_{kl}
- \frac{\partial^2}{\partial_l \partial \tau}h_{jm}
\frac{\partial^2}{\partial_l\partial_i}h_{km} +
\frac{\partial^2}{\partial \tau^2}h_{jl}
\frac{\partial^2}{\partial_i
\partial \tau}h_{lk} \right) \la{rr2}. \ee

Note that in Eq. (\re{Fou1}) there are two creation (annihilation)
operators, one for each polarization state, with the usual
commutation relations: \bn \left[a(p,{\bf k}), a^{\dagger}(p',{\bf
k'})\right] &=& \delta_{pp'}
\delta^3({\bf k} - {\bf k'}),\nonumber \\
\left[a(p,{\bf k}), a(p',{\bf k'})\right] &=& 0, \nonumber \\
\left[a^{\dagger}(p,{\bf k}), a^{\dagger}(p',{\bf k'})\right] &=& 0.
\en The presence of the creation-annihilation operators and their
commutation relations was not considered by Alexander {\it et. al.}
\ci{SPS}, and that led them to conclusions very different from ours
about the vacuum expectation value of the operator $R \tilde R$ as
we will show.

Note also that the polarization tensors behave similarly to the
vector polarizations for the electromagnetic field: the polarization
tensor $\epsilon_{ij}(p,{\bf k})$ is ``orthogonal" to the
propagation direction {\bf k} of the gravitational wave: \be k_i
\epsilon_{ij}(p,{\bf k}) = 0. \la{dot} \ee In addition, the
polarization tensors are ``orthonormal": \be
\epsilon^\ast_{ij}(p,{\bf k})\epsilon_{ij} (p',{\bf k}) =
2\delta_{pp'}. \ee This, of course, is equivalent to say that the
``cross" product between two different polarization tensors results
in a unitary vector in the direction of propagation of the
gravitational waves: \bn \epsilon^{ilk} \epsilon_{ij}^{\ast}(+, {\bf
k}) \epsilon_{jl}(\times, {\bf k}) &=& -\epsilon^{ilk}
\epsilon_{ij}^{\ast}(\times, {\bf k}) \epsilon_{jl}(+, {\bf
k}) = 2\frac{k_k}{\left|{\bf k}\right|}, \nonumber \\
\epsilon^{ilk} \epsilon_{ij}^{\ast}(+, {\bf k}) \epsilon_{jl}(+,
{\bf k}) &=& \epsilon^{ilk} \epsilon_{ij}^{\ast}(\times, {\bf k})
\epsilon_{jl}(\times, {\bf k}) = 0. \la{pro} \en We have proved the
expressions in Eq. (\re{pro}) by using properties of the rotational
transformations.

Thus, the vacuum expectation value $<0 \mid  R \tilde R \mid 0 >$
can now be calculated: \be \langle 0 \mid R \tilde R \mid 0 \rangle
= -\frac{2}{a^4} \, \epsilon^{ijk} \left\langle 0 \left|
\frac{\partial^2}{\partial_l \partial \tau}h_{jm}
\frac{\partial^2}{\partial_m \partial_i}h_{kl} -
\frac{\partial^2}{\partial_l \partial \tau}h_{jm}
\frac{\partial^2}{\partial_l\partial_i}h_{km} +
\frac{\partial^2}{\partial \tau^2}h_{jl}
\frac{\partial^2}{\partial_i
\partial \tau}h_{lk} \right| 0 \right\rangle \la{rr2}. \ee

Using the properties of the polarization tensors, and the
commutation relations of the creation-annihilation operators it is
easy to see how each of the three contribution in Eq. (\re{rr2})
vanishes. The contributions from the second and third term vanish as
the bracket $\langle 0 \mid ... \mid 0 \rangle$ implies the same
polarization state for the creation-annihilation operators involved.
This in turn means that we are left with the products
$\epsilon^{ijk} \epsilon_{jm}^{\ast}(p, {\bf k})\epsilon_{km}(p,
{\bf k})$ which, according to the Eq. (\re{pro}), are zero. Finally,
the contribution from the first term vanishes as the spacial
derivatives involved lead to products of the form $k_i
\epsilon_{ij}(p,{\bf k})$ which are, of course, zero (see Eq.
(\re{dot})).

As a conclusion we can say that, contrary to what was claimed by
Alexander {\it et. al.} \ci{SPS} who found a quite positive
result, the matter-antimatter asymmetry cannot be generated from
the contribution of the tensor cosmological perturbations to the
gravitational anomaly in the Standard Model.

\medskip

{\bf Acknowledgments}

Y.R. wants to acknowledge Lancaster University and Universities UK
for their partial financial help and the Colombian agencies
COLCIENCIAS and COLFUTURO for their postgraduate scholarships. C.Q.
thanks Lancaster University by its hospitality and PPARC by its
financial support during his visit to Lancaster University.

\renewcommand{\refname}{{\large References}}

\end{document}